\documentclass[useAMS,usenatbib,usegraphicx]{mn2e}

\def\m@th{\mathsurround=0pt }

\def\eqalign#1{\null\,\vcenter{\openup1\jot\m@th

 \ialign{\strut\hfil$\displaystyle{##}$&$\displaystyle{{}##}$\hfil

 \crcr#1\crcr}}\,}

\title[2dFGRS: Butcher-Oemler effect] {The 2dF Galaxy Redshift Survey: The blue 
galaxy fraction and implications for the Butcher-Oemler effect}
\author[Roberto De Propris et al.]{
\parbox[t]{\textwidth}{
Roberto De Propris$^3$\thanks{propris@mso.anu.edu.au},
Matthew Colless$^{1,3}$,
John A.\ Peacock$^{12}$,
Warrick Couch$^5$,
Simon P.\ Driver$^3$,
Michael L. Balogh$^2$,
Ivan K.\ Baldry$^9$,
Carlton M.\ Baugh$^2$,
Joss Bland-Hawthorn$^1$,
Terry Bridges$^{1,17}$, 
Russell Cannon$^1$, 
Shaun Cole$^2$, 
Chris Collins$^4$, 
Nicholas Cross$^9$,
Gavin Dalton$^{6,15}$, 
George Efstathiou$^7$, 
Richard S.\ Ellis$^8$, 
Carlos S.\ Frenk$^2$, 
Karl Glazebrook$^9$, 
Edward Hawkins$^{11}$,
Carole Jackson$^{16}$,
Ofer Lahav$^7$, 
Ian Lewis$^6$, 
Stuart Lumsden$^{10}$, 
Steve Maddox$^{11}$,
Darren Madgwick$^{13}$,
Peder Norberg$^{14}$,
Will Percival$^{12}$,
Bruce A.\ Peterson$^3$, 
Will Sutherland$^{12}$,
Keith Taylor$^8$ (the 2dFGRS Team)}
\vspace*{6pt} \\ 
$^1$Anglo-Australian Observatory, P.O.\ Box 296, Epping, NSW 2111,
    Australia\\  
$^2$Department of Physics, University of Durham, South Road, 
    Durham DH1 3LE, UK \\ 
$^3$Research School of Astronomy \& Astrophysics, The Australian 
    National University, Weston Creek, ACT 2611, Australia \\
$^4$Astrophysics Research Institute, Liverpool John Moores University,  
    Twelve Quays House, Birkenhead, L14 1LD, UK \\
$^5$Department of Astrophysics, University of New South Wales, Sydney, 
    NSW 2052, Australia \\
$^6$Department of Physics, University of Oxford, Keble Road, 
    Oxford OX1 3RH, UK \\
$^7$Institute of Astronomy, University of Cambridge, Madingley Road,
    Cambridge CB3 0HA, UK \\
$^8$Department of Astronomy, California Institute of Technology, 
    Pasadena, CA 91025, USA \\
$^9$Department of Physics \& Astronomy, Johns Hopkins University,
       Baltimore, MD 21118-2686, USA \\
$^{10}$Department of Physics, University of Leeds, Woodhouse Lane,
       Leeds, LS2 9JT, UK \\
$^{11}$School of Physics \& Astronomy, University of Nottingham,
       Nottingham NG7 2RD, UK \\
$^{12}$Institute for Astronomy, University of Edinburgh, Royal Observatory, 
       Blackford Hill, Edinburgh EH9 3HJ, UK \\
$^{13}$Lawrence Berkeley National Laboratory, 1 Cyclotron Road,
       Berkeley, CA 94720, USA \\ 
$^{14}$ETHZ Institut fur Astronomie, HPF G3.1, ETH Honggerberg, CH-8093
       Zurich, Switzerland \\
$^{15}$Rutherford Appleton Laboratory, Chilton, Didcot, OX11 0QX, UK \\
$^{16}$CSIRO Australia Telescope National Facility, PO Box 76, Epping,
       NSW 1710, Australia \\
$^{17}$Physics Department, Queen's University, Kingston, ON, K7L 3N6,
       Canada \\
}

\begin{document}

\date{}


\maketitle


\begin{abstract}

We derive the fraction of blue galaxies in a sample of clusters at $z < 0.11$ and the
general field at the same redshift. The value of the blue fraction is observed to
depend on the luminosity limit adopted, cluster-centric radius and, more generally,
local galaxy density, but it does not depend on cluster properties. Changes in the
blue fraction are due to variations in the relative proportions of red and blue
galaxies but the star formation rate for these two galaxy groups remains 
unchanged. Our results are most consistent with a model where the star formation
rate declines rapidly and the blue galaxies tend to be dwarfs and do not favour
mechanisms where the Butcher-Oemler effect is caused by processes specific to 
the cluster environment.

\end{abstract}


\section{Introduction}

The observation by \cite{bo84} that clusters of galaxies contain a larger
fraction of blue galaxies at progressively higher redshift (the Butcher-Oemler
effect) contributed to the establishment of the current view of clusters as sites 
of active galaxy evolution, in which environmental influences dramatically alter
the morphologies and star formation histories of their members.

However, as the number of clusters observed has increased, it has become 
clear that the Butcher-Oemler effect is not solely or simply an evolutionary trend. 
The large scatter observed in the blue fractions for clusters in narrow redshift ranges 
\citep{sma98,mar00,got03} implies the existence of environmental effects, which may 
compete with, and possibly mimic, evolutionary trends, if the cluster samples evolve 
with redshift.

Studies of the original Butcher-Oemler sample lend support to this scenario: \cite{nkb88} 
measured the surface densities and velocity dispersions of seven Butcher-Oemler clusters 
and found that the high and low redshift sample differ. \cite{ae99} show that the X-ray 
luminosity of  high redshift Butcher-Oemler clusters is higher than that of low redshift 
clusters, although the X-ray luminosity  function of clusters does not evolve to $z \sim 0.8$
\citep{hol02}, arguing that the high redshift sample does not represent the progenitors of 
modern-day clusters. Similarly, both \cite{sma98} and \cite{fai02} find a low blue fraction 
in their samples of X-ray selected clusters.

The blue fraction has also been observed to depend on a number of other factors,
such as:  the luminosity limit used and the cluster centric distance \citep{ell01,got03}, 
richness \citep{mar01}, cluster concentration \citep{bo84} and, possibly, the presence of 
substructure \citep{mru00}. These findings point to the necessity of understanding environmental 
effects on the blue fraction in order to disentangle evolutionary trends from selection biases 
in studies of clusters at high redshift.

The aim of the present paper is to study how the fraction of blue galaxies varies as
a function of cluster properties and in the field at the same redshift in the local universe, 
in order to analyse the effects of environment in isolation from the evolutionary trends that 
give rise to the Butcher-Oemler effect. \cite{lew02} and \cite{bal03} have recently discussed 
how the star formation rate, as measured from the H$\alpha$ equivalent width, varies with
environment within the 2dFGRS sample. However, colours provide a measure of the
average star formation histories over longer timescales than H$\alpha$ and may therefore
better reflect the influence of environment on galaxy properties, especially if the mechanisms
responsible for the Butcher-Oemler effect operate on longer time frames, as is the case for
harassment \citep{moo96}, for instance.

The structure of this paper is as follows: in the next section we present our
analysis; selection of clusters, cluster members and a comparison field sample, 
definition of the radial and luminosity limits used, $k$-corrections, colour-magnitude 
relations, density measurements and a discussion of how the blue fraction and its
error are measured. The main results are detailed in \S 3. below and we discuss these 
in \S 4.  Throughout this paper we adopt a cosmology with $\Omega_M=0.3$, $\Omega_{
\Lambda}=0.7$. It is normal to define $h \equiv H_0 / 100\, {\rm km\ s^{-1}\ Mpc^{-1}}$:
here we suppress the $h$ scaling, so that the full explicit meaning of Mpc in length
units is $h^{-1}$ Mpc; for absolute magnitudes $M$ stands for $M+5 \log_{10} h$.

\section{Analysis}

In this work, we analyse how the blue fraction of galaxies varies as a function of
cluster properties and field density at $z < 0.11$; our purpose is to provide a 
local reference for studies of galaxy colour evolution in high redshift clusters
(the Butcher-Oemler effect). It is therefore useful to reproduce, in our analysis,
some of the features of the original \cite{bo84} study, as this is the basis for
most current work on the evolution of stellar populations of cluster galaxies.

The original analysis of \cite{bo84} defined blue galaxies as being (i) within a radius 
containing 30\% of the cluster population; (ii) brighter than a no-evolution $k-$corrected 
$M_V=-20$ and (iii) bluer by 0.2 magnitudes in $B-V$ (no-evolution and $k-$corrected) 
than the colour-magnitude relation defined by the cluster early-type galaxies. Below we 
describe how we  implemented these prescriptions for our dataset.

\subsection{Cluster and member selection}

The clusters analyzed here are those studied by \cite{dep03} in their paper on the
composite galaxy luminosity function. This sample consists of 60 clusters at $z < 0.11$
containing at least 40 spectroscopic members and with average 85\% completeness.
Although these clusters share, to some extent, the biases of the Abell, APM and EDCC
catalogues from which they were originally drawn (see De Propris et al. 2002 for
details), these 60 objects provide an at least approximately complete and volume
limited ensemble of nearby clusters, spanning a large range of properties (such as
richness and velocity dispersion) determined from the same data \citep{dep02}.

Cluster membership was determined via a `double gapping' method, as described in
greater detail in \cite{dep02}: for each putative cluster centre and redshift, we first
required that the likely cluster members be surrounded by 1000 km s$^{-1}$ gaps
on either side of the redshift distribution; we next computed a first estimate for the 
mean $cz$ and velocity dispersion ($\sigma_r$) and ranked all galaxies in order of 
distance from the mean $cz$. If galaxies had a `gap' in redshift space from their neighbour 
larger than our first estimate of $\sigma_r$ all galaxies with a faster or slower velocity were 
excluded.  This procedure yields a sample of cluster members which is likely to be relatively
uncontaminated by interlopers (i.e. field galaxies with appropriate redshift but not
dynamically bound to the cluster).

\subsection{Aperture and luminosity selection}

We calculate $r_{30}$ (the original aperture used by Butcher \& Oemler 1984) by using 
all cluster members within a 3 Mpc radius from the central galaxy. We then choose an
aperture than includes 30\% of these galaxies. It is observed that the value of $r_{30}$ 
varies depending on cluster dynamics and central concentration. For this reason it may be
generally more appropriate to use apertures based on cluster dynamical properties, in order 
to insure that all galaxies considered share the same environment in all clusters. \cite{ell01} 
advocate using $r_{200}$, the radius at which the cluster is 200 times denser than the general
field \citep{cye97}, or its multiples. This is defined as:

$$r_{200}={{\sqrt{3} \sigma_r} \over {10 H(z)}} \eqno(1) $$

\noindent where $H(z)$ represents the redshift-dependent Hubble constant. In the following, we 
will use both $r_{200}$ and $r_{200}/2$ for our analysis. 

The original limit by \cite{bo84} corresponds to approximately 1.8 magnitudes below the 
$M^*$ point. Here, we use a magnitude limit 1.5 magnitudes fainter relative to our computed 
$M^*$ and we also analyse the effects of using fainter limits. This allows us to study a more 
consistent sample of galaxies in all clusters.\\

\subsection{$k$-corrections}

$B_{\rm J}-R_{\rm F}$ colours for all 2dFGRS galaxies were derived using data from the 
SuperCosmos survey \citep{him01}. Given the colours and redshifts we derived a $k$-correction 
in the following way. 

The {\sc kcorrect} package \citep{bla03a} fits a combination of
templates to galaxy colour data, and thus yields a consistent $k$-correction.
Although only two bands are available for the 2dFGRS, this approach was validated
by fitting the full DR1 $u$$g$$r$$i$$z$ data, then comparing with the result
of fitting $gr$ only. The differences in K-correction are generally
at the 0.01 mag. level.

The main problem with {\sc kcorrect} is that the templates are unable to
describe extremely red galaxies ($B_J-R_F=1.26$ corresponding to $g-r=0.87$), 
and so a consistent $k$-correction cannot be obtained. In such cases, we took 
an alternative approach, based on the models of \cite{bc03}. Each galaxy was 
modelled by a single burst, varying the age until the observed colour was 
matched at the given redshift. This age is degenerate with metallicity; in 
practice, we assumed 0.4 times Solar metallicity ($Z=0.008$), except for very 
red galaxies where a current age of $>13$~Gyr was required. In these cases, 
the current age was set at $13$~Gyr and the metallicity increased until the colour was 
matched. For red galaxies, $k$-corrections deduced in this way agree almost exactly with
the results of {\sc kcorrect}, and we were able to match smoothly from one to
the other to deal with the very red galaxies that {\sc kcorrect} cannot model.
The results differ for extremely blue galaxies (at the 0.1 mag. level), but a
single burst is a rather unrealistic model in such cases and the results
of {\sc kcorrect} are to be preferred where they are consistent.

The fitting functions given below describe the $k$-correction results for
all colours of practical interest at $z<0.3$, to within a maximum error
of about 0.02 mag. Given the intrinsic uncertainties in both photometric
modelling and calibration, and in the interests of clarity, these residuals
were ignored and the fit was treated as exact.

$$
\eqalign{
k(B_{\rm J}) = \, &(-1.63+4.53 x) y -  (4.03+2.01 x)  y^2  \cr
&- {z\over1+(10 z)^4} \cr
}
\eqno(2)$$

$$ k(R_{\rm F})=(-0.08+1.45 x)  y - (2.88+0.48 x) y^2 \eqno(3)$$

\noindent where

$$ x=B_{\rm J}-R_{\rm F} \eqno(4)$$

$$ y= z/(1+z) \eqno(5) $$

\subsection{Colour-magnitude relations}

For all clusters, we determine the colour-magnitude relation by carrying out a least absolute 
deviation regression fit to the observed colour distribution \citep{ak78}. The average observed 
slope is about $-0.019 \pm 0.003$, consistent with the relation observed in the Coma galaxies 
for $V-R$ (Eisenhardt et al. 2004, in preparation). There appears to be some cluster-to-cluster
variation in the slope, but this may be due to small number statistics in some cases.

We carry out the same analysis for field galaxies, but we divide the sample according to local 
density (see below for details). The slope for the field galaxies is $\sim -0.013 \pm 0.001$.
These slopes were calculated robustly and jackknife resampling was then used to compute the errors 
on the values of the slopes. The slopes of the relations for field and cluster galaxies appear to 
be different at about the $2 \sigma$ level, but we do not regard this as compelling (cf., Hogg et 
al. 2004 for a similar study from Sloan colours).

The 2dFGRS SuperCosmos colours can be transformed to the SDSS $g$ and $r$ bands using the 
following equations:

$$ B_{\rm J}=g+0.15+0.13 (g-r) \eqno(6) $$

$$ R_{\rm F}=r-0.13 \eqno(7) $$

\noindent A difference in $B-V$ of 0.2 magnitudes from the ridge line defined by early-type 
galaxies is equivalent to a 0.2 magnitude difference in $g-r$ \citep{got03}. From the above 
equation we then derive that the corresponding difference in $B_{\rm J} - R_{\rm F}$ is 
0.23 magnitudes. Fig.~1 shows the colour-magnitude relation and the colour distribution, 
marginalised over the derived colour magnitude relation, for a few representative clusters. 

\begin{figure*}
\label{fig1}
\centering\includegraphics[width=150mm]{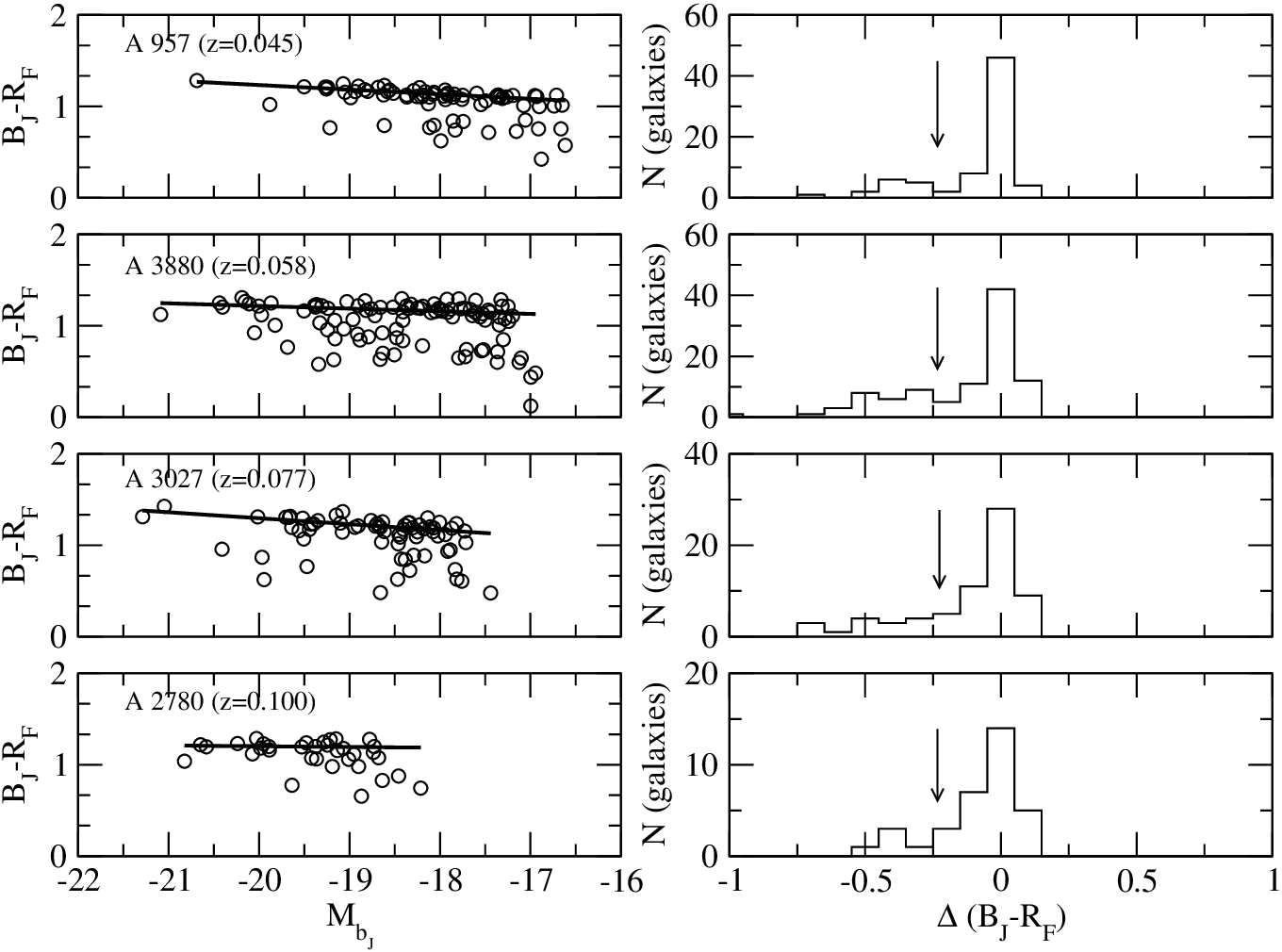}
\caption{Colour-magnitude relation (left) and histogram (right) of marginalised colour 
distribution (opposite each colour-magnitude relation) for four representative clusters. 
The arrow in the histogram indicates the blue/red separation (see text)}
\end{figure*}

\subsection{Calculation of local density}

For field galaxies, the only relevant `environmental' property is their local density. 
We calculate this by computing the number of galaxies to $M_{B_{\rm J}}=-19$ in an 8 
Mpc sphere centred on each galaxy, with appropriate completeness corrections. A detailed 
description of this procedure is given in \cite{cro04}. Unfortunately, this assumes that
all radial velocities are due to the smooth Hubble flow and can be used as proxies for distance.
This is not applicable in the cluster environment, where the galaxies have turned around
and no longer participate in the cosmological expansion (see, e.g., Cooperstock et al. 1998).
For clusters, therefore, we assume that all members are effectively within $r_{200}$ and
calculate the density in an 8 Mpc sphere containing only cluster members. This
places the density for cluster galaxies on the same scale as that used for field
galaxies. As we note below, this is likely to be a slight underestimate of the
actual density (as some members may exist beyond the virial radius).

\subsection{Calculation of the blue fraction}

Calculation of the error in the derived blue fraction has been carried out with a variety 
of recipes. Here we derive the appropriate formulation for our sample of cluster members. 
Note that if the numbers of red and blue galaxies are determined via background subtraction, 
extra contributions to the error statistics due to clustering need to be included.

The blue fraction is defined as the ratio of $m$ blue galaxies observed out of $n$ total galaxies. 
Assuming that $m$ and $n$ obey Poisson statistics, the blue fraction is:

$$f_b= m/n \eqno(8)$$

\noindent and its likelihood has the same form independent of whether we assume Poisson or
binomial statistics (i.e. with $n$ fixed in advance):

$$ L  \propto f_b^m (1-f_b)^{n-m} \eqno(9)$$

\noindent whose maximum is, trivially, $m/n$. Therefore the variance of
the blue fraction is:

$$\sigma^2 (f_b)= \bigg({{d^2 \ln L} \over {d f^2_b}}\bigg)^{-1}=
{{m (n-m)} \over n^3} \eqno(10)$$

\noindent This is incorrect for $m=0$ because $dL/df \not= 0$ for $m=0$. 
In that  case we ask what $f_b$ yields likelihood equal to $\exp (-1/2\,
f_{max})$. This is approximately $1/2n$, which is a reasonable error bar 
to adopt for the $m=0$ case.

In our analysis we only use spectroscopically confirmed cluster members.
Since the redshift sample is $B_{\rm J}$ selected, without regard to colour,
this should not bias the result, especially given our high completeness
(cf. Ellingson et al. 2001 for a similar approach to data in the CNOC2
survey).

\section{Results}

\subsection{Luminosity and radial dependence}

The blue fraction depends on the luminosity limit and on the aperture used to include 
members \citep{mar00,ell01}. We analyse these dependencies here, using our sample.
Fig.~2 shows the blue fraction, calculated to $M^*+1.5$, vs. $z$ for all three radii we 
consider: $r_{30}$, r$_{200}$ and r$_{200}/2$ (where $M^*$ is taken from De Propris et
al. 2003a). This plot shows that there is no dependence of the blue fraction on redshift 
and that therefore our sample of objects is adequate to study environmental trends in 
isolation from the effects of evolution. Unless otherwise specified, we refer to this 
blue fraction as $f_b$ for the remainder of this paper.

\begin{figure}
\centering\includegraphics[width=80mm]{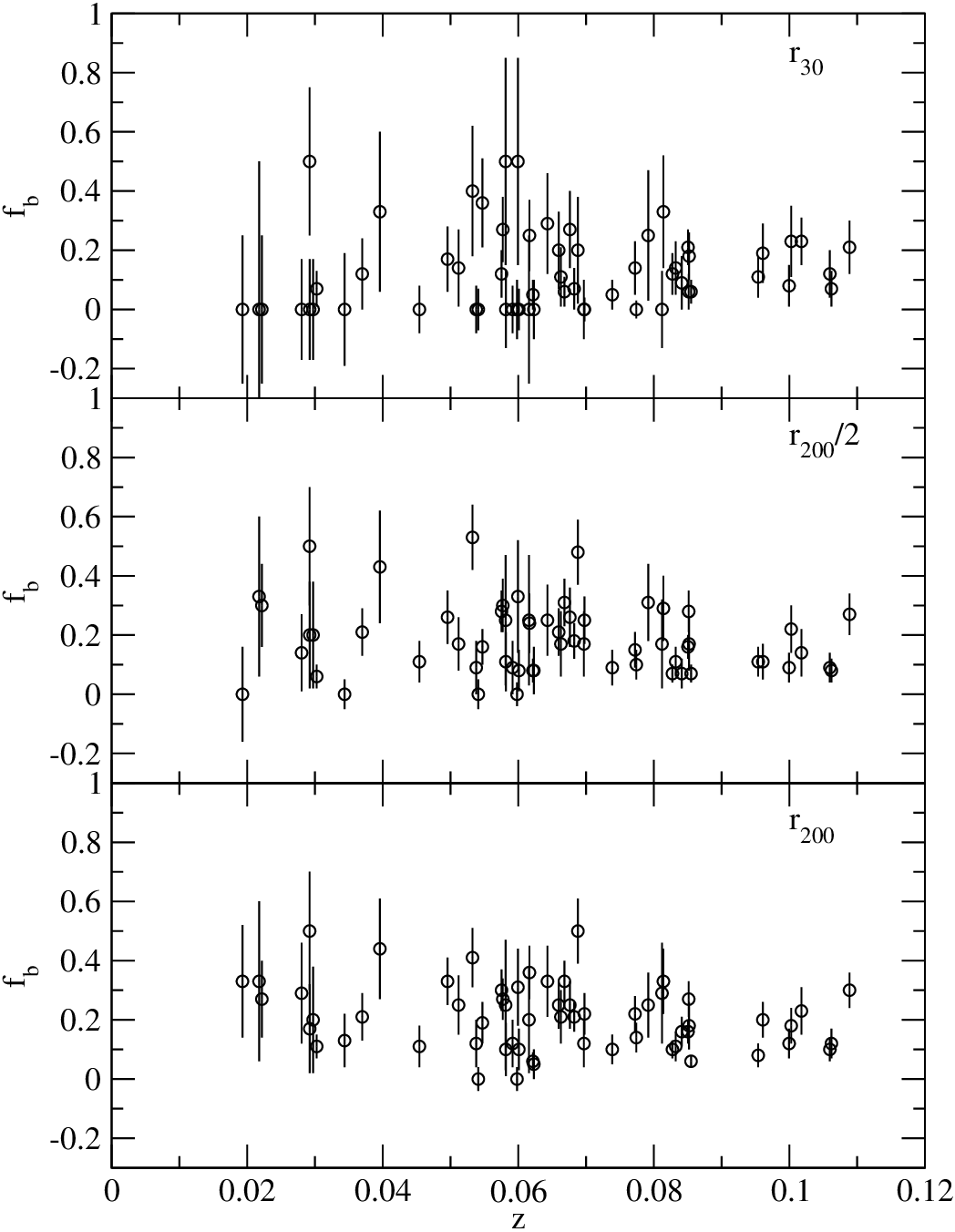}
\caption{ 
The blue fraction to $M^*+1.5$ for all clusters, in all three apertures
considered}
\end{figure}

Inspection of Fig.~2 shows that there is considerable scatter in the derived blue fraction. 
This is consistent with previous studies of local and X-ray selected samples \citep{sma98,
mar01,got03}. We compare the distribution of points with a Gaussian distribution with mean and
standard deviation derived from a weighted fit to the data. The distribution shows significant 
skewness and a $\chi^2$ test yields probabilities below 10$^{-3}$ that the data points are 
consistent with a single value of $f_b$. This suggests that the scatter observed is a real 
property of the sample.

We plot the mean values of f$_b$ as a function of the magnitude limit used for all three 
apertures in Fig.~3. A $t-$test shows that the distributions have significantly different 
means: the blue fraction increases as a function of radius and as a function of luminosity. 
This is consistent with previous findings by \cite{mar01} and \cite{ell01} and implies that 
the the blue galaxies are intrinsically faint objects and reside preferentially in the cluster 
outskirts.

\begin{figure}
\centering\includegraphics[width=80mm]{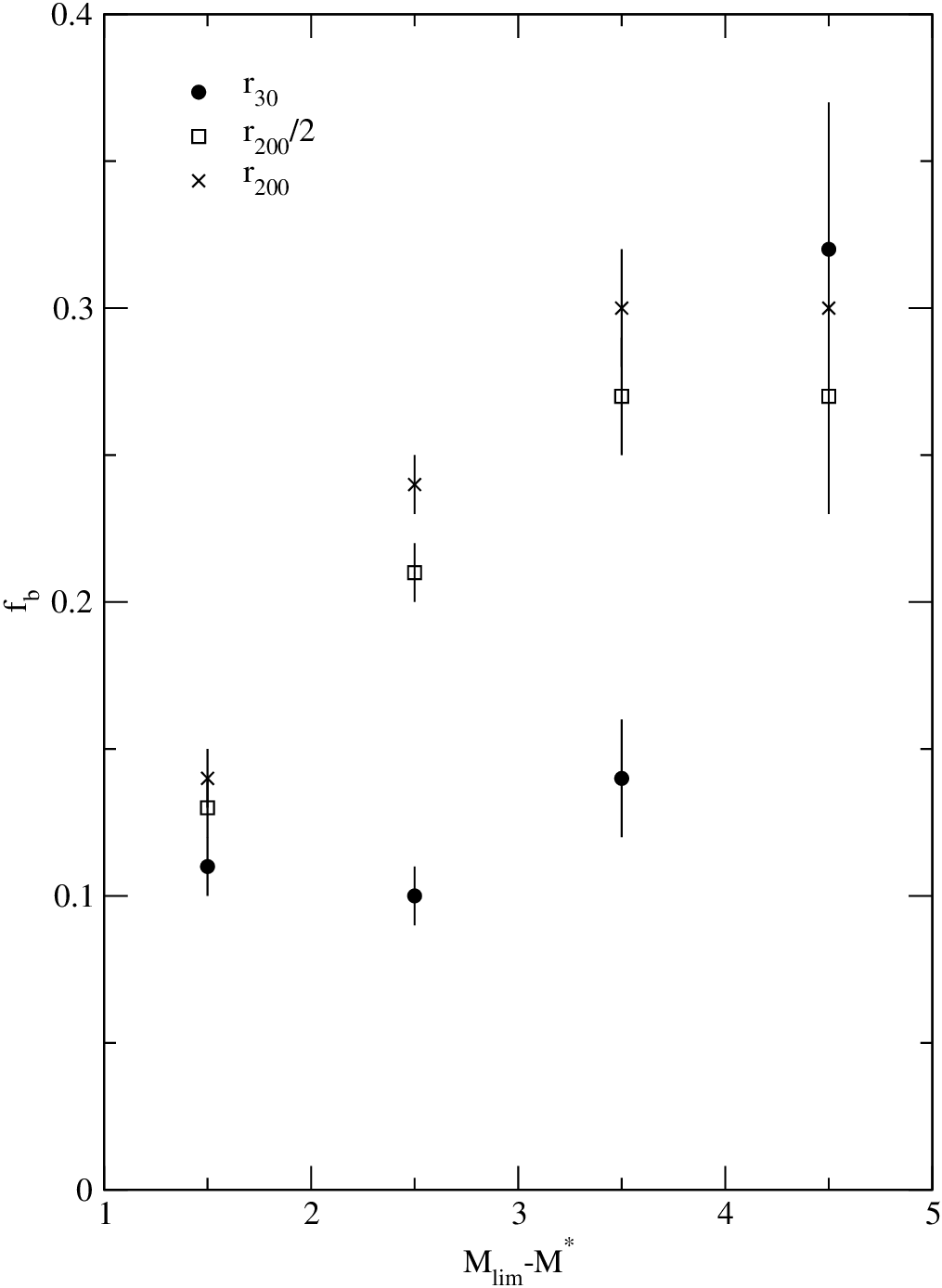}
\caption{ 
Mean values of the blue fraction as a function of luminosity and aperture}
\end{figure}

\subsection{Dependence on cluster properties}

We will next consider how the blue fraction depends on cluster properties derived from our 
original study \citep{dep02}: Bautz-Morgan type, richness, presence of substructure, 
velocity dispersion and concentration.

The Bautz-Morgan type measures the dominance of the brightest cluster galaxy relative 
to the rest of the cluster population. It is used as an indicator of dynamical evolution, 
where the brightest galaxy evolves via the merger of $M^*$ galaxies at the cluster center. 
Velocity dispersion provides a measure of the cluster mass (assuming the clusters to be 
virialized) and of the relative speed of galaxy encounters (which is important for some 
of the mechanisms for the origin of the Butcher-Oemler effect). Richness, as measured 
from the number of galaxies brighter than $M_{b_{\rm J}}=-19$ \citep{dep03}, is an 
indicator of cluster mass and mean density. We consider the probability that a 
cluster contains substructure, using the Lee-Fitchett statistics \citep{fit88}. Although 
there is no metric to determine the amount of substructure a cluster contains, the probability 
of its being significantly substructured provides a measure of the amount of recent merging 
a cluster has undergone and therefore may allow us to estimate how recently the cluster 
has formed. Finally, we consider the dependence on cluster concentration, defined as

$$C=\log (r_{60}/r_{20}) \eqno(11)$$

\noindent where $r_{20}$ and $r_{60}$ are the radii containing 20\% and 60\% of 
the cluster population, respectively.

Fig.~4 shows how the blue fraction (to $M^*+1.5$) depends on these five variables. 
We only plot the case corresponding to $r_{200}/2$ for economy of space, but the 
results are similar in all three apertures. In no case is there any clear evidence of 
any dependence of the blue fraction on Bautz-Morgan type, richness, velocity dispersion,
probability of substructure or concentration. This is confirmed by a non-parametric 
Spearman test, which returns low correlation probabilities in all instances.

\begin{figure}
\centering\includegraphics[width=80mm]{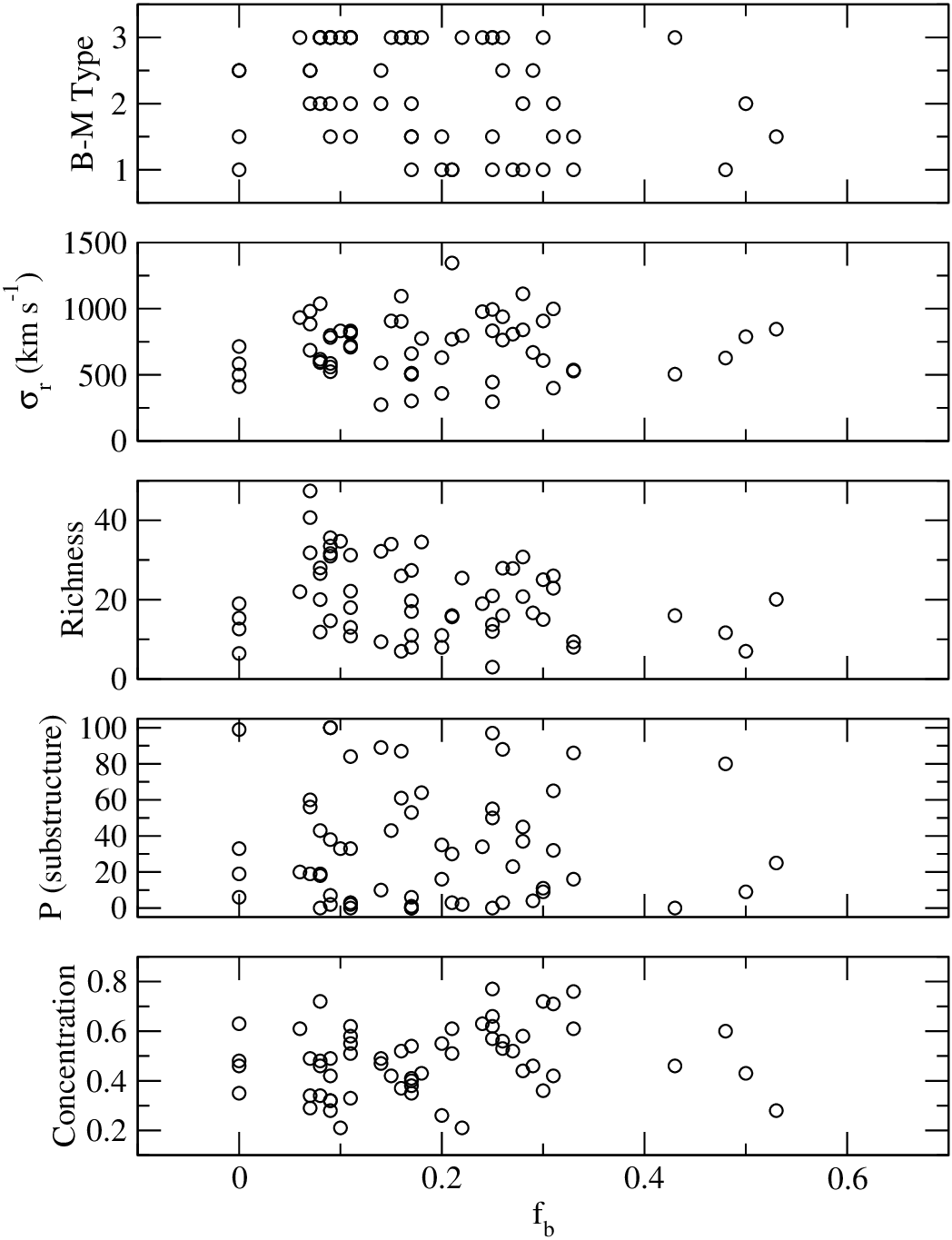}
\caption{ 
Dependence of the measured blue fraction on cluster properties for
the $r_{200}/2$ aperture
}
\vskip 5mm
\end{figure}

\begin{figure}
\centering\includegraphics[width=75mm]{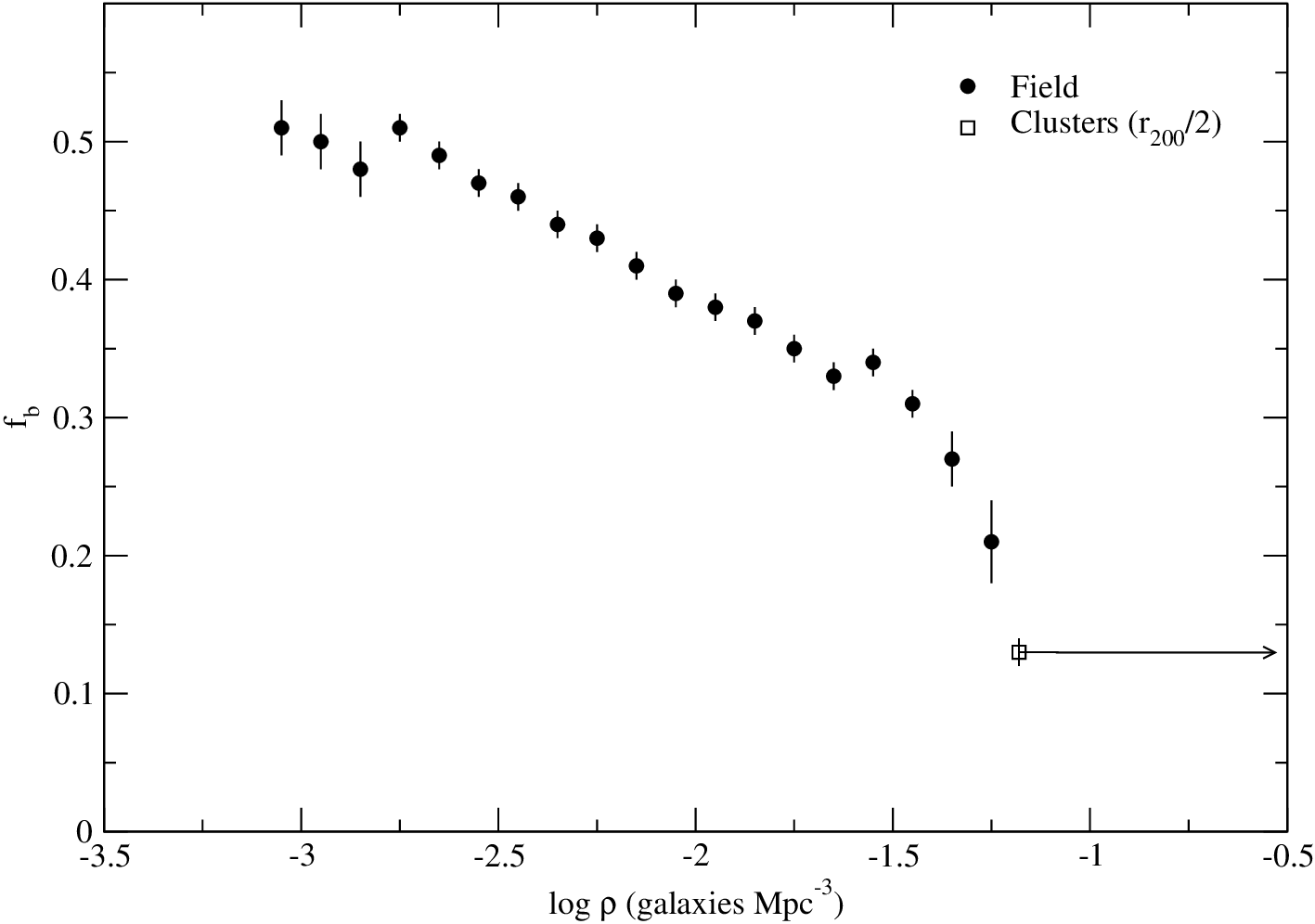}
\caption{ 
The blue fraction in the general field as a function of local density
(filled circles) and the mean for all clusters within $r_{200}/2$ (open
square). Density is defined within 8 Mpc spheres as in Croton et al. (2004).
For clusters we use the number of galaxies within $r_{200}$ and assume
that all cluster members are within this sphere in our calculation of
density. Because this is likely to be a slight underestimate, we indicate
in the figure that this density is only a lower limit.}
\vskip 5mm
\end{figure}

\subsection{Density dependence in the field}

We have selected a sample of field galaxies from the 2dFGRS galaxy redshift survey with the 
same redshift distribution as the clusters. For each galaxy we derived the local density in 
an 8 Mpc sphere (Croton et al. 2004). From this sample we determine the blue fraction as 
a function of local density in Fig.~5.

The blue fraction follows a well defined linear trend, becoming higher in low density regions. 
This trend can be extended to clusters, whose behaviour is then simply the more extreme
version of the relation between $f_b$ and local density as observed in the general field.

Galaxies then display a bimodal distribution in colour space, with well-defined red and 
blue wodges, whose relative populations change with density. This represents a confirmation
of the trends observed by the SDSS in their more extensive photometry \citep{hogg03,
bla03}.

It is interesting to compare how the relative star formation rates, as measured from 
H$\alpha$ equivalent widths, vary with density or with the only properties that, in clusters, 
appear to affect the blue fraction, luminosity and radius (computing absolute star formation
rates is difficult with 2dF data, as the spectra are not easy to flux calibrate, but H$\alpha$
equivalent widths can be used to provide an estimate of the relative star formation rate within
the sample). The H$\alpha$ equivalent widths were computed using Gaussian line fitting with a 
small ($2$ \AA) correction for underlying line absorption. No dust extinction correction was 
carried out. A full description of the procedures used can be found in \cite{lew02}, especially 
their section 2.4.

The variation of H$\alpha$ equivalent width with density, luminosity and radius is shown in 
Fig.~6 and 7. for red and blue galaxies in both environments. We see that, while the blue 
fraction changes the mean relative star formation rate does not. The bimodal  behaviour 
observed in the colours extends to the H$\alpha$ equivalent width, as noticed by \cite{bal03}. 
The changes in blue fraction and H$\alpha$ equivalent width as a function of density are 
due to changes in the relative fractions of quiescent and star forming galaxies.

\begin{figure}
\centering\includegraphics[width=75mm]{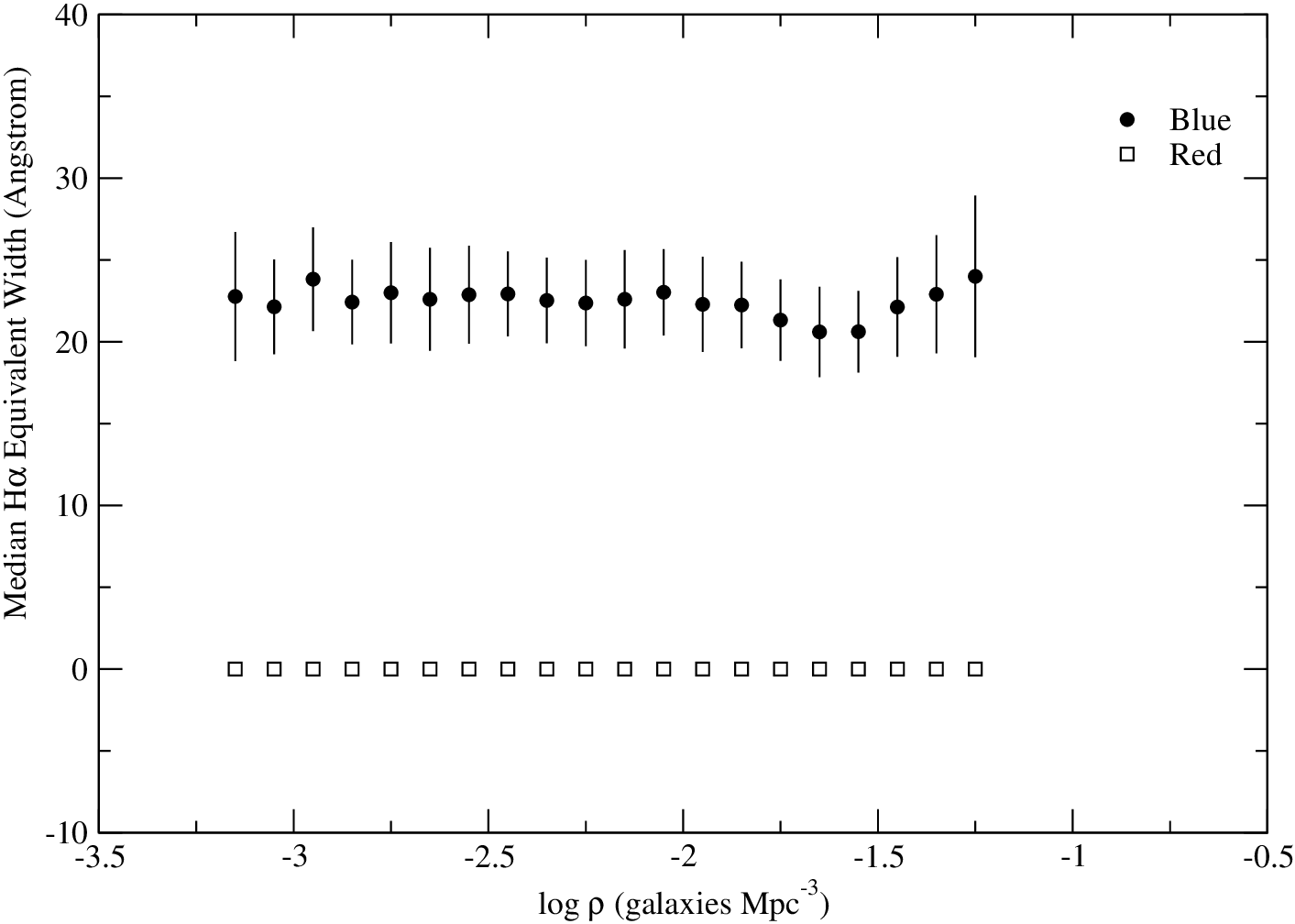}
\caption{ 
Median H$\alpha$ equivalent widths (in Angstroms) for blue and red
galaxies in the field as a function of density. Errors are interquartile
means. The errors for the red galaxies are smaller than the symbols.
}
\vskip 10mm
\end{figure}

\begin{figure}
\centering\includegraphics[width=80mm]{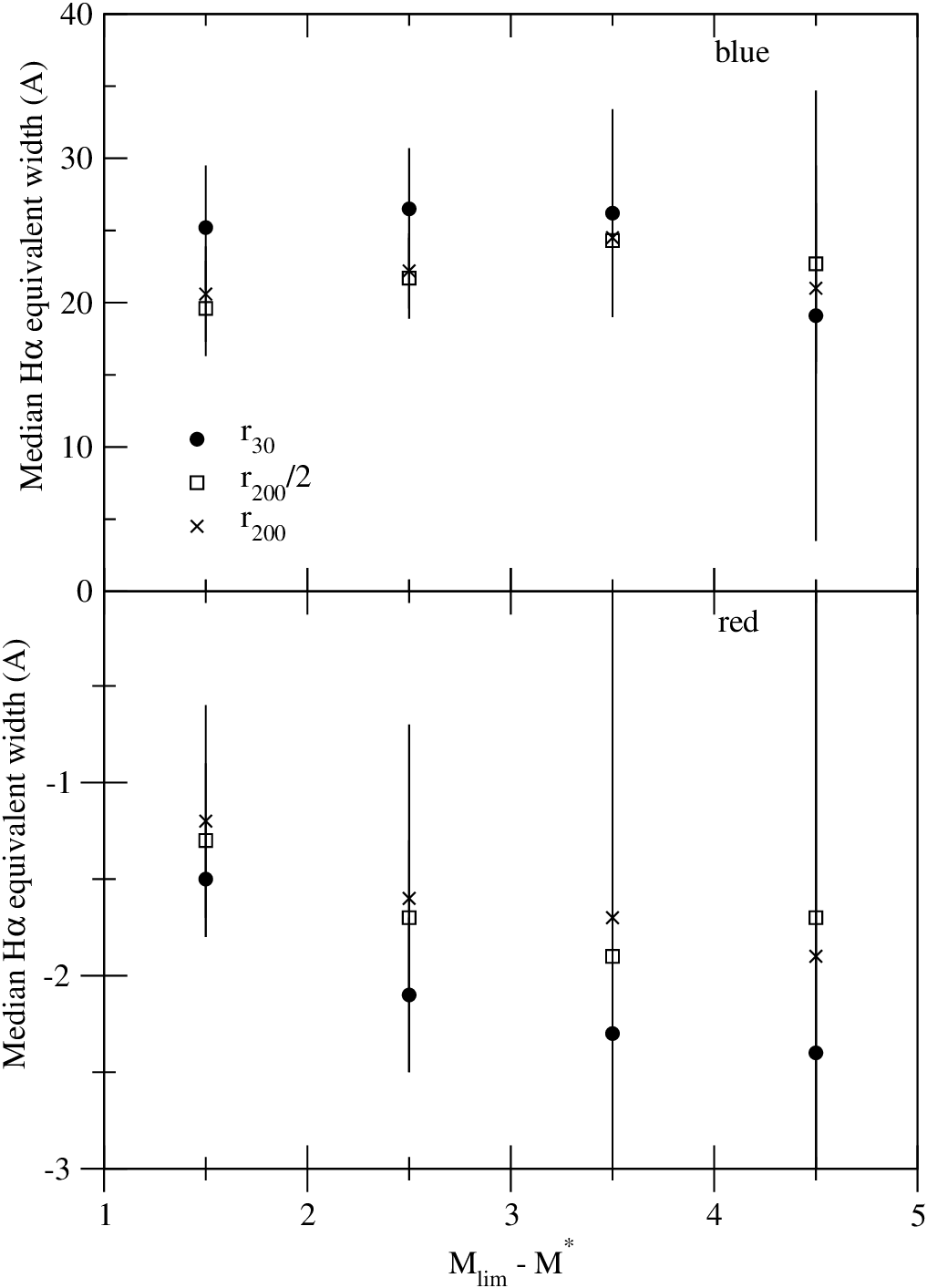}
\caption{ 
Median H$\alpha$ equivalent widths (as in Fig.~6) for blue and red
galaxies in clusters as a function of luminosity and radius. Errors are
interquartile means.}
\end{figure}

\section{Discussion}

We have determined the fraction of blue galaxies in a sample of nearby clusters and
in the general field at the same redshift. Although the blue fraction varies considerably
from cluster to cluster, we find that the variation is a real property of the sample and
not due to statistical noise.

The mean value of the blue fraction is $0.11 \pm 0.01$ for galaxies within $r_{30}$ and
$0.13 \pm 0.01$ for galaxies within $r_{200}/2$. These are higher than the value of
$0.03 \pm 0.09$ from \cite{mar00}, $\sim 0.07$ from \cite{mar01} and $\sim 0.07$ from
\cite{pim02} at $z < 0.11$, although they are within the errors. One possibility for this
discrepancy lies in the possibility of blue interlopers from the field population being
included in our redshift sample \citep{dia01}. We have calculated the level of contamination
by integrating the field luminosity function of \cite{mad02} over the appropriate luminosity
range and within the pseudo-volume defined by the aperture used and the range in
velocities spanned by cluster galaxies. The blue fraction for this sample was calculated
from our density-dependent blue fraction. The average level of contamination
is $0.08 \pm 0.06$ but this depends critically on assumptions concerning the mean density
of galaxies in cluster outskirts (which influence both the normalization and the fraction
of blue interlopers as per Fig.~5). In addition, there are contributions to the error from both
Poisson and clustering statistics. The number of blue galaxies is, in any case, small and
therefore the level of contamination is uncertain and the error in its estimate quite
large. We therefore ignore the issue of possible contamination of our spectroscopic sample 
of cluster members.

\begin{itemize}

\item A first conclusion to be drawn from the derived blue fractions is that few clusters have 
$f_b > 40\%$ and none has $f_b > 60\%$ .

Since this sample of clusters is, at least to first order, complete and volume limited, it 
represents a fair sampling of the range of blue fractions encountered in cluster environments. 
By contrast, samples of clusters at high redshift contain at least a few objects with blue 
fractions in excess of 40\% (Fairley et al. 2002, La Barbera et al. 2003 and references
therein). This would suggest that the observed evolution is real: contamination in our sample 
(if real) should only make the local blue  fraction lower, while high redshift clusters are 
drawn from the high richness envelope that contains few local clusters with high $f_b$. However, 
it is possible that optically selected clusters at high redshift tend to contain higher fractions 
of blue galaxies because these make them more conspicuous in blue plates.\\

\item The blue fraction appears to depend on the luminosity limit and cluster centric radius 
used (Fig.~3), as previously found by \cite{mar00} and \cite{ell01}.  

This behaviour is similar to that observed for dwarf galaxies in Coma and Abell 2218: \cite{pra03} 
have shown that dwarfs have a steeper luminosity function at larger cluster centric radii and 
are preferentially found at large distances from the cluster core, with these trends being 
stronger for the faintest dwarfs. Similar trends may have been observed for galaxies in the Coma 
cluster \citep{bej02} and appear to persist in the $z\sim 0.4$ CNOC sample. To the extent that 
present-day clusters and their populations represent proxies for high redshift objects, these 
observations imply that a proportion of the blue galaxies are intrinsically low luminosity 
objects.\\

\item There is no dependence of the blue fraction on cluster properties (Fig.~4). The lack of
correlation of the blue fraction with cluster properties suggests that the star formation rate,
or its decrease, is not related to the large scale structure: therefore, mechanisms which
involve cluster-wide effects, such as ram stripping by the cluster gas, tides, which depend
on the cluster mass, or harassment, whose efficiency is proportional to velocity dispersion,
are disfavoured by the present study, as well as other explanations that depend on processes
specific to the cluster environment. However, as \cite{bal03} show, the star formation rate
is affected by local processes and close interactions are a viable mechanism for the origin of
the blue fraction. 

The above is apparently in contrast  with some previous studies: \cite{mru00} claim that clusters 
containing large amounts of substructure have higher blue fractions and interpret this in terms 
of a shock model during collisions of subclusters. However, their result is based on three 
clusters and may not be valid for the general population. \cite{mar01} and \cite{got03} suggest 
that the blue fraction depends on richness. One possibility is that, since they use a fixed 0.7 
Mpc aperture, and given the radial dependence observed in Fig.~3, that a spurious richness 
dependence may be induced by aperture effects. We have derived blue fractions for a 0.7 Mpc 
aperture in our clusters and tested for correlation with richness but failed to detect a 
significant signal.  Similarly, we failed to detect any correlation with richness if we use 
the same luminosity range as \cite{mar01}. However, we see in Fig.~4 that rich clusters tend 
not to have large blue fractions and we also observe that the blue fraction in the 0.7 Mpc 
aperture shows large scatter. If the sample used by \cite{mar01} and \cite{got03} is biased 
towards richer clusters at high redshift, the combination of these two effects may produce a 
spurious correlation.\\

\item We observe that the blue fraction exhibits a strong dependence with local density in 
the general field (Fig.~5) and that the cluster value represents a continuation of this trend 
to higher density regimes.

This suggests that the same processes are responsible for the Butcher-Oemler effect in all 
environments and that these mechanisms vary smoothly as a function of density. Again, this
implies that cluster-specific processes are not likely causes of changes in the blue fraction,
but local mechanisms whose efficiency varies smoothly with density, such as interactions 
\citep{lh88,cou98}, may be viable explanations.\\

\item The relative star formation rate for the blue galaxies does not appear to decrease as 
a function of density (Fig.~6) or as a function of luminosity or radius in clusters (Fig.~7).

This implies that changes in the blue fraction are solely due to changes in the relative 
fractions of red (quiescent) and blue (star-forming) galaxies. The trends we observe can
be roughly reproduced by using the density-dependent luminosity functions of 
\cite{cro04} for early- and late-type galaxies and the type-dependent cluster luminosity 
functions of \cite{dep03} and simply assuming that early-type galaxies are red and late types 
are blue. This suggests that the simple model presented in \cite{dep03}, where galaxies 
simply moved between spectral types without number or luminosity evolution, provides, 
heuristically, a good representation of galaxy evolution.\\

One possible caveat is that, by selecting blue galaxies, we have automatically selected for
galaxies with H$\alpha$ emission, while morphologically selected samples show declining star
formation rates as a function of density \citep{gom03,got03b}. However, the observation that
blue cluster galaxies have the same H$\alpha$ equivalent width as their counterparts in the
field is non-trivial, because lower (but non-zero) star formation rates will lead to weaker
H$\alpha$ emission, while the colour would remain blue by our definition. This is in contrast,
for instance, with models where galaxies are `choked' in clusters \citep{bal00}.

\end{itemize}

The above argues for a model where star formation declines over relatively short timescales, 
leading to the bimodal distribution in H$\alpha$ observed by \cite{bal03}, and galaxy colours
evolve quickly to the red envelope, producing the colour bimodality observed here and
in the SDSS \citep{hogg03,bla03}. As in \cite{dep03} this is possible if the optical
colour is dominated by the young population but the majority of light is provided by
the underlying old stars, so that once the star formation is extinguished galaxy
colours quickly evolve on to the passive locus. \cite{shi02} have shown that truncating 
the star formation of spiral galaxies in the field leads them on to the
colour-magnitude relation defined by the Coma E/S0 galaxies, but that this process
is inefficient for the more massive galaxies and is viable only for low mass spirals.
When we consider the radial and luminosity trends, one possible interpretation is
that a large fraction of the blue galaxies are actually dwarfs undergoing episodes of
star formation, as also suggested by a number of other lines of evidence \citep{ros97,
cou98,dep04}. However, we caution the reader that since our sample of cluster and 
field galaxies is local, we cannot properly discuss the origin of the blue fraction
at higher redshift. 

A complication in interpreting these data is that the observed correlations of $f_b$ and
H$\alpha$ with density on large scales imply that the environment {\it today} is not
affecting the properties of the population and therefore we are unable to observe
galaxy evolution in progress but just its end result. This implies that identifying the
mechanisms responsible for the Butcher-Oemler effect in the local universe is problematic.
Surveys at intermediate redshifts (e.g. DEEP2, VIMOS) may be able to witness the main phases 
of galaxy evolution.

\section*{Acknowledgements}

We are indebted to the staff at the Anglo-Australian Observatory
for their tireless effort and assistance in supporting 2dF during
the course of the survey. We are also grateful to the Australian and
UK time assignment committees for their continued support for this
project. The work of Warrick J. Couch is supported by a grant from
the Australian Research Council. Michael L. Balogh acknowledges PPARC
fellowship PPA/P/S/2001/00298. We wish to thank the referee, Tomotsugu
Goto, for an useful report, which greatly improved the clarity of this
paper.

\setlength{\bibhang}{2.0em}

\clearpage

\end{document}